\documentclass[conference]{IEEEtran}
\IEEEoverridecommandlockouts

\usepackage{cite}
\usepackage{amsmath,amssymb,amsfonts}
\usepackage{algorithmic}
\usepackage{graphicx}
\usepackage{textcomp}
\usepackage{xcolor}
\usepackage{booktabs}
\usepackage{lipsum}
\usepackage{soul}
\usepackage{multirow}
\usepackage{caption}
\usepackage{subcaption}
\usepackage{siunitx}
\usepackage{url}

\def\BibTeX{{\rm B\kern-.05em{\sc i\kern-.025em b}\kern-.08em
    T\kern-.1667em\lower.7ex\hbox{E}\kern-.125emX}}

\begin{document}
\bstctlcite{IEEEexample:BSTcontrol} 

\title{Guiding WaveMamba with Frequency Maps for Image Debanding
\thanks{This work has been supported by the UKRI MyWorld Strength in Places Programme (SIPF00006/1).}
}

\author{Xinyi Wang, Smaranda Tasmoc, Nantheera Anantrasirichai, and Angeliki Katsenou\\
School of Computer Science, University of Bristol, Bristol BS1 8UB, UK}

\maketitle

\begin{abstract}
Compression at low bitrates in modern codecs often introduces banding artifacts, especially in smooth regions such as skies. These artifacts degrade visual quality and are common in user-generated content due to repeated transcoding. We propose a banding restoration method that employs the Wavelet State Space Model and a frequency masking map to preserve high-frequency details.  Furthermore, we provide a benchmark of open-source banding restoration methods and evaluate their performance on two public banding image datasets. Experimentation on the available datasets suggests that the proposed post-processing approach effectively suppresses banding compared to the state-of-the-art method (a DBI value of 0.082 on BAND-2k) while preserving image textures. Visual inspections of the results confirm this. Code and supplementary material are available at: \url{https://github.com/xinyiW915/Debanding-PCS2025}.
\end{abstract}


\section{Introduction}
\label{sec: intro}
Despite the outstanding performance of recent image and video coding standards, such as Versatile Video Coding (VVC)~\cite{VVC}, Alliance for Open Media Video 1 (AV1)~\cite{AV1}, and High Efficiency Video Coding (HEVC)/H.265~\cite{HEVC}, streaming content live and/or at low bitrates requires aggressive compression, which often introduces compression artifacts that compromise visual quality. Banding is a type of quantization artifact that appears when a low-texture region of an image is encoded with insufficient bit depth. It is typically more noticeable in plain, smooth backgrounds, such as skylines, and becomes visible along a diagonal or curved edge when one color gradation shifts to another, creating a staircase-like effect. Banding is often encountered in user-generated content, as multiple rounds of transcoding on sharing platforms can further amplify the artifacts. 

Various techniques are designed for banding and contour artifacts in images and video based on the point of intervention, either before compression~\cite{daly2003bit,10.1145/383259.383326,1057702} or within an image/video encoder~\cite{casali2015adaptive,yoo2009loop} (known as in-loop processing). Here, we focus on \textit{post-processing} methods because they are flexible and codec-independent. These methods either detect banded regions and intervene locally~\cite{choi2006false,yoo2009loop,xu2024adaptive}, or directly enhance the entire decoded image~\cite{bhagavathy2009multiscale}.

Post-processing methods can be further categorized into \textit{traditional image processing}  techniques and \textit{deep learning based} solutions. Examples of the first category include the following. Choi et al.~\cite{choi2006false} proposed a method combining directional dilation and edge-preserving filtering, where directional dilation enhances true edges while reducing false contours, and filtering protects image details from blurring. Yoo et al.~\cite{yoo2009loop} addressed contouring through an acyclic method that adds pseudo-random noise to blocks containing false contours, although it struggles with subtle artifacts. Bhagavathy et al.~\cite{bhagavathy2009multiscale} introduced multiscale probabilistic dithering, which processes images at multiple scales using Gaussian pyramids and applies dithering at each level to mitigate contouring, though this can introduce noise and increase computational demands. Jin et al.~\cite{jin2011composite} proposed a block-based detection approach using a composite model that balances gradient cues and blocks edge smoothness to eliminate false contours. Tu et al.~\cite{tu2020adaptive} developed AdaDeband, an adaptive method with three key components: a banding detector, a content-aware filter that reconstructs gradients while preserving textures, and a dithering step that improves visual quality by re-quantizing smooth regions. Last, FFmpeg~\cite{ffmpeg} provides two practical filters: Gradfun, which uses filtering and dithering on gradient areas to reduce banding, and deband, which focuses on removing pronounced banding while preserving detail. Both are less effective for heavily quantized inputs. 

Recent examples of deep learning-based solutions for banding include the deepDeband model~\cite{zhou2022deep}, which employs a conditional Generative Adversarial Network (GAN) and is available in two variants: deepDeband-F, which is faster but less precise, and deepDeband-W, which adopts a bilateral weighting strategy to eliminate block discontinuity. In the same category of generative approaches falls cGAN~\cite{chang2025two} that uses a U-Net to first segment the image into a flat area mask, which then guides the generative network. Finally, the authors in~\cite{xu2024adaptive} propose a method in the frequency domain that effectively isolates and eliminates components associated with banding artifacts, employing partitioned dynamic convolution for spatial differentiation of banded and non-banded features.


In this paper, we present a debanding framework based on the alignment of a joint spatial-frequency representation with the human visual system. The frequency domain not only accurately captures the overall activity level of the luma component in an image, but also intuitively quantifies other perceptual features, such as contrast information. In banded areas, we observe medium to high frequencies in smooth areas, which opposes the pristine version. Based on this mechanism, we adopt a spatial frequency masking strategy. To this end, we employ the Wavelet State Space Model (WaveMamba)~\cite{zou2024wave} and explore different strategies to create weighted wavelet maps, combining the benefits of both banded and restored images: preserving true edges, structures, and textures, while removing banding artifacts. In summary, this paper offers the following contributions:
\begin{itemize}
    \item We provide a benchmark of the latest open-sourced methods in banding restoration, along with a set of image restoration methods that have been successful with other types of distortions.
    \item We propose a post-processing restoration technique that performs frequency-based mapping to guide a WaveMamba model into successfully restoring the banded images.
\end{itemize}
The rest of the paper is organized as follows. Section~\ref{sec: method} presents the WaveMamba network and frequency masking. Section~\ref{sec: exp} details the datasets and benchmarking setup and results are reported in Section~\ref{sec: res}. Section~\ref{sec: con} concludes the paper.

\section{Proposed Method}
\label{sec: method}
We employ the WaveMamba~\cite{zou2024wave} architecture designed for low-light image enhancement and improve it by introducing a weighted wavelet map mechanism. 

\subsection{WaveMamba Network}
WaveMamba~\cite{zou2024wave} combines Discrete Wavelet Transform (DWT) and state-space modeling to capture low- and high-frequency areas, enabling simultaneous brightness and texture restoration. As banding artifacts mainly occur in low-frequency regions (smooth areas) and high-frequency details still need to be preserved in non-banded areas, WaveMamba can effectively separate low- and high-frequency information across multiple scales, suppressing artifacts and restoring textures.

The input image $I_{\text{in}} \in \mathbb{R}^{H \times W \times 3}$ is first processed through a convolutional network to generate an initial feature map $F_{\text{in}} \in \mathbb{R}^{H \times W \times C}$. Subsequently, it undergoes a three-level DWT decomposition, where each level downsamples the spatial resolution by a factor of 2:
\begin{equation}
\label{eq: dwt}
    \begin{aligned}
    &\{F_L^{i}, F_H^{(i,h)}, F_H^{(i,v)}, F_H^{(i,d)}\} = \text{DWT}(F_L^{i-1}), \\
    &F_L^{i} \in \mathbb{R}^{\frac{H}{2^i} \times \frac{W}{2^i} \times C}, i=1,2,3,\quad F_L^{0} := F_{in}.
    \end{aligned}
\end{equation}
Here, $F_L^{i}$ denotes the low-frequency component, while $F_H^{(i,h)}$, $F_H^{(i,v)}$, and $F_H^{(i,d)}$ represent the horizontal, vertical, and diagonal high-frequency components, respectively.

The low-frequency feature $F_L^{i}$ is passed through a Low-Frequency State Space Block (LFSSB), which employs a state space model to capture long-range dependencies. The computation is defined as follows:
\begin{equation}
F_L^{i+1} = \mathcal{G}(\mathcal{S}(\text{LN}(F_L^{i})) + \beta \cdot F_L^{i}) + \gamma \cdot (\mathcal{S}(\text{LN}(F_L^{i})) + \beta \cdot F_L^{i}),
\end{equation}
where $\mathcal{S}(\cdot)$ denotes the Vision State Space Module, $\mathrm{LN}(\cdot)$ represents layer normalization, $\mathcal{G}(\cdot)$ is the Gated Feed Forward Network, and $\beta$ and $\gamma$ are learnable scaling factors. The LFSSB captures low-frequency details from the spatial domain. The high-frequency component $F_H^{i}$ is reduced in the channel dimension using Selective Kernel Feature Fusion, then combined with the low-frequency component $F_L^{i+1}$ and fed into the High-Frequency Enhance Block (HFEB). This block uses the low-frequency feature to guide the restoration of high-frequency details and consists of two parts:
\begin{equation}
\label{eq: hfeb}
    \begin{aligned}
    \tilde{F}_H &= \mathcal{M}(\mathrm{LN}(F_H^{i}), F_L^{i+1}) + F_H^{i} \\
    F_H^{(i,E)} &= \mathcal{C}(\mathrm{LN}(\tilde{F}_H), F_L^{i+1}) + \tilde{F}_H,
    \end{aligned}
\end{equation}
where $\mathcal{M}(\cdot, \cdot)$ denotes the Frequency Matching Transformation Attention, $\mathcal{C}(\cdot, \cdot)$ denotes the Frequency Correction Feedforward Network, $\mathrm{LN}(\cdot)$ represents layer normalization, and $F_H^{(i,E)}$ is the restored high-frequency feature. The HFEB enhances high-frequency components by selecting similarity features from low-frequency components.

The overall WaveMamba network is constructed using a multi-scale U-Net~\cite{ronneberger2015u} structure. After feature extraction and enhancement through the LFSSB and HFEB, the decoder reconstructs features in reverse. At each scale, high-frequency and low-frequency components are fused and refined using the same blocks, then upsampled via the Inverse Discrete Wavelet Transform (IDWT):
\begin{equation}
F_{\text{out}} = \text{IDWT}(F_L^{i+1}, F_H^{(i,E)}).
\end{equation}
This process is repeated across all levels, progressively restoring spatial resolution, until the upsampled features are fed into the same convolution layer to obtain the intermediate output.

\subsection{Frequency Masking Map}
\label{sec: masking}
To effectively remove banding artifacts while preserving structural details, it is crucial to distinguish flat areas (often affected by banding) from high-frequency areas such as edges and textures. Directly enhancing the banded input can often lead to over-smoothing or detail loss. To address this, we introduce a frequency masking strategy using a Weighted Wavelet Map (WWM), $M_w$. The masking map selectively preserves high-frequency components while obtaining debanded areas.  We explore the best strategy of $M_w$ through the following three variants.

\noindent
\textbf{1. WaveMamba-WWM}:
$M_w$ is computed in the wavelet domain from the input image and applied as a post-processing step after prediction. $I_{\text{in}}$ is converted to grayscale, and a multi-level DWT is applied. From the final decomposition, horizontal, vertical, and diagonal detail coefficients $(c_h, c_v, c_d)$ are extracted. A wavelet magnitude map $S = |c_h| + |c_v| + |c_d|$ is computed, and then normalized to [0, 1] to form $M_w$:
\begin{equation}
M_w = \frac{S - \min(S)}{\max(S) - \min(S)}.
\end{equation}
$M_w$ assigns different weights to the banded input $I_{\text{in}}$ and debanded output $I_{\text{out}}$, preserving details in textured regions while suppressing artifacts in flat areas:
\begin{equation}
I_{\text{deband}} = M_w \cdot I_{\text{in}} + (1 - M_w) \cdot I_{\text{out}}.
\end{equation}

\noindent
\textbf{2. WaveMamba-DWT}: 
In this variant, frequency masking is integrated into the WaveMamba model during training. As the encoder performs multi-level DWT decomposition (Eq.~\ref{eq: dwt}), we utilize the high-frequency components $(F_H^{(i,h)}, F_H^{(i,v)}, F_H^{(i,d)})$ from each level to construct the masking map. At each scale $i$, we compute the absolute sum of the components, apply channel-wise averaging, and perform bilinear interpolation to upsample to the original resolution. $M_w$ is obtained by averaging across all scales, followed by normalization:
\begin{equation}
    S_i = \text{Interp} \left( \text{Mean} \left( |F_H^{(i,h)}| + |F_H^{(i,v)}| + |F_H^{(i,d)}| \right) \right),
\end{equation}
\begin{equation}
    M_w = \text{Norm} \left( \frac{1}{L} \sum_{i=1}^{L} S_i \right).
\end{equation}
Here, $i$ denotes the DWT level index, and $L = 3$ is the total number of decomposition levels. Unlike WWM, WaveMamba-DWT enhances debanded output by leveraging multi-scale wavelet weights within the encoder, enabling end-to-end learning of frequency-aware fusion.

\noindent
\textbf{3. WaveMamba-MAP}: 
Here, the masking mechanism is also incorporated during training. $M_w$ is derived from the HFEB outputs, computed after enhanced high-frequency components (Eq.~\ref{eq: hfeb}). These features $F_H^{(i,E)}$ from each level are aggregated, upsampled, and processed via channel-wise averaging and normalization:
\begin{equation}
    S_i = \text{Interp} \left( \text{Mean} \left( |F_H^{(i,E)}| \right) \right),
\end{equation}
\begin{equation}
    M_w = \text{Norm} \left( \frac{1}{L} \sum_{i=1}^{L} S_i \right).
\end{equation}
Compared to WaveMamba-DWT, WaveMamba-MAP relies on features refined through learning, making the $M_w$ more adaptive to high-frequency structural context. Both variants involve retraining the model with the learned mask embedded in the forward pass.

\section{Experimental Design}
\label{sec: exp}
\subsection{Datasets}
There are not many public dataset available on debanding. The deepDeband dataset~\cite{zhou2022deep} consists of 1,439 pairs of ground truth and quantized full HD images (1920×1080) with annotated banded and non-banded regions, from which overlapping 256×256 patches were extracted. Only patches containing banding, along with their corresponding pristine counterparts, were retained, yielding 51,420 pairs of image patches. The dataset spans diverse visual content and is of sufficient scale to support the development of deep learning models. We trained our WaveMamba variants on deepDeband using a 70-20-10 split: 35,994 patch pairs for training, 10,284 for validation, and 5,142 for testing.

The BAND-2k~\cite{chen2024band} is the largest image dataset for evaluating banding artifacts, containing 2,000 distorted images generated using different compression and quantization methods. These images were sourced from over 870 diverse videos spanning computer graphics, user-generated, and professional content. BAND-2k includes subjective scores, along with patch-level annotations of banded and non-banded regions in quantized images. However, it is unsuitable for training models due to the lack of paired pristine and distorted image patches. Therefore, we use BAND-2k only for cross-dataset evaluation and inference in our experiments.

\subsection{Evaluation Metrics}
Objective quality metrics are essential for assessing the perceptual quality of enhanced images compared to their pristine counterparts. While widely adopted artifact-specific metrics such as Peak Signal-to-Noise Ratio (PSNR), Structural Similarity Index (SSIM)~\cite{wang2004image}, Learned Perceptual Image Patch Similarity (LPIPS)~\cite{zhang2018unreasonable}, and Video Multi-Method Assessment Fusion (VMAF)~\cite{li2016toward} have proven effective for general distortions, they fall short in detecting banding artifacts due to the subtle and localized nature of false contours. 

To address this limitation, dedicated banding detection metrics such as Blind BANding Detector (BBAND)~\cite{tu2020bband}, Contrast-aware Multiscale Banding Index (CAMBI)~\cite{tandon2021cambi}, and Deep Banding Index (DBI)~\cite{kapoor2021capturing} have been proposed to quantify banding specifically. 
BBAND~\cite{tu2020bband} is a video quality metric that estimates banding severity directly from the input. It computes a frame-level score based on the band visibility map and local gradient statistics, without requiring ground truth. 
CAMBI~\cite{tandon2021cambi} improves perceptual alignment by incorporating the contrast sensitivity function~\cite{seshadrinathan2009image} to capture human visual sensitivity.
DBI~\cite{kapoor2021capturing} quantifies banding artifacts through a no-reference deep neural network model that produces both an overall banding score and a spatial banding map, accurately reflecting the severity and distribution of banding. Banding-specific metrics are expected to align better with human perception, although they may be less reliable under resolution changes or in non-banded regions.


\subsection{Benchmarking Methods}
To assess banding restoration, we benchmarked traditional and deep learning-based methods, ranging from classical signal processing filters to deep generative models.
The Gradfun filter~\cite{ffmpeg} in FFmpeg is a traditional technique for mitigating banding artifacts via gradient interpolation and dithering. It smooths pixel transitions in flat regions but fails to remove false contours in textured areas.
The Plug-and-Play Priors (PPP)~\cite{venkatakrishnan2013plug} framework incorporates learned denoising models as implicit priors, embedded within an optimization based on the alternating direction method of multipliers.  

Deep Image Prior (DIP)~\cite{ulyanov2018deep} uses the implicit bias of convolutional networks as a handcrafted prior. An initialized network is gradually optimized to reconstruct a single corrupted image without external training data. For debanding, we use a randomly initialized U-Net to suppress banding artifacts during reconstruction.
The Latent Diffusion Model (LDM)~\cite{rombach2022high} is a generative model that operates in a perceptual latent space. It compresses an input image into a latent representation using a trained encoder, applies a diffusion process in the latent space, and reconstructs the image with a decoder. We adapt LDM for image super-resolution, treating banding removal as an enhancement of a degraded input.
The deepDeband~\cite{zhou2022deep} is a GAN-based model built on the Pix2Pix framework, featuring a generator that produces debanded outputs and a discriminator that ensures realism. We include only deepDeband-F for full-image processing as the inference model in the performance comparison.
To evaluate the effectiveness of the proposed frequency masking map, we integrate it into both deepDeband and our WaveMamba variants. We enhance deepDeband-F by fusing its output with the banded input via WWM, resulting in deepDeband-F-WWM, which helps suppress over-smoothing introduced by the generator. In WaveMamba variants, frequency masking is used to mitigate banding in smooth regions while enhancing fidelity in detailed areas during reconstruction.

\section{Results and Discussion}
\label{sec: res}
\subsection{Performance Comparison}

\begin{table}[t]
    \centering
    \scriptsize
    \setlength{\tabcolsep}{4.1pt} 
    \caption{Quantitative comparison on the deepDeband~\cite{zhou2022deep} test set (patch pairs). Best and second-best results are highlighted in \textbf{bold} and \underline{underline}, respectively. Arrows indicate if higher or lower values are preferred.}
    \label{tab: comparison}
    \begin{tabular}{@{}lccccccc@{}}
    \toprule
    \textbf{Model} & \textbf{PSNR$\uparrow$} & \textbf{SSIM$\uparrow$} & \textbf{LPIPS$\downarrow$} & \textbf{CAMBI$\downarrow$} & \textbf{BBAND$\downarrow$} & \textbf{DBI$\downarrow$} \\
    \midrule
    Banded Images & 36.286 & 0.977 & 0.081 & 1.467 & 1.217 & 0.083 \\
    \midrule
    FFmpeg~\cite{ffmpeg} & 31.383 & 0.941 & 0.110 & 0.955 & 0.632 & 0.035 \\
    PPP~\cite{venkatakrishnan2013plug} & 32.432 & 0.927 & 0.188 & 3.619 & 1.230 & \textbf{0.005} \\
    \midrule
    DIP~\cite{ulyanov2018deep} & 34.364 & 0.957 & 0.093 & 1.728 & 0.820 & 0.034 \\
    LDM~\cite{rombach2022high} & 30.597 & 0.905 & 0.140 & 0.824 & 0.853 & 0.057 \\
    deepDeband-F~\cite{zhou2022deep} & 34.584 & 0.964 & 0.071 & \underline{0.057} & \textbf{0.464} & 0.122 \\
    deepDeband-F-WWM & 35.005 & 0.967 & 0.066 & \textbf{0.055} & \underline{0.484} & 0.104 \\
    \midrule
    WaveMamba~\cite{zou2024wave} & \textbf{42.399} & \textbf{0.989} & 0.037 & 3.947 & 1.410 & \underline{0.021} \\
    WaveMamba-WWM & \underline{42.201} & \underline{0.988} & \textbf{0.033} & 3.615 & 1.436 & \underline{0.021} \\
    WaveMamba-DWT & 42.111 & \underline{0.988} & \underline{0.036} & 3.798 & 1.403 & 0.022 \\
    WaveMamba-MAP & 41.546 & \underline{0.988} & 0.037 & 3.523 & 1.343 & 0.022 \\
    \bottomrule
    \end{tabular}
\end{table}

\begin{table}[t]
    \centering
    \scriptsize
    \setlength{\tabcolsep}{4.1pt} 
    \caption{Cross-dataset evaluation of WaveMamba variants on BAND-2k~\cite{chen2024band} dataset, using models trained on deepDeband~\cite{zhou2022deep}.}
    \label{tab: cross_dataset}
    \begin{tabular}{@{}lcccccc@{}}
        \toprule
        \textbf{Model} & \textbf{PSNR$\uparrow$} & \textbf{SSIM$\uparrow$} & \textbf{LPIPS$\downarrow$} & \textbf{CAMBI$\downarrow$} & \textbf{BBAND$\downarrow$} & \textbf{DBI$\downarrow$} \\
        \midrule
        Banded Images & 37.344 & 0.964 & 0.100 & \textbf{0.513} & 0.564 & 0.431 \\
        \midrule
        deepDeband-F~\cite{zhou2022deep} & 33.130 & 0.897 & 0.116 & 0.561 & \textbf{0.198} & 0.225 \\
        deepDeband-F-WWM & 33.263 & 0.898 & 0.115 & \underline{0.557} & \textbf{0.198} & 0.227 \\
        \midrule
        WaveMamba~\cite{zou2024wave} & 38.926 & \textbf{0.977} & 0.063 & 0.647 & 0.519 & \underline{0.148} \\
        WaveMamba-WWM & \textbf{39.054} & \textbf{0.977} & \textbf{0.059} & 0.639 & 0.541 & 0.157 \\
        WaveMamba-DWT & \underline{38.963} & \textbf{0.977} & \underline{0.062} & 0.702 & 0.551 & 0.158 \\
        WaveMamba-MAP & 38.505 & \underline{0.976} & 0.066 & 0.691 & \underline{0.476} & \textbf{0.082} \\
        \bottomrule
    \end{tabular}
\end{table}

Table~\ref{tab: comparison} reports the quantitative comparison on the deepDeband test set. While metrics like CAMBI and BBAND are designed to assess banding artifacts, the results show that reduced banding does not lead to better structural consistency. For example, deepDeband-F and deepDeband-F-WWM achieve the lowest BBAND score (0.464) and the lowest CAMBI score (0.055), respectively, yet fail to improve PSNR or SSIM, indicating a lack of structural fidelity. In contrast, WaveMamba achieves the highest PSNR (42.399dB) but has the highest CAMBI and BBAND scores, indicating the worst banding artifacts, which do not align with visual inspection. Notably, the proposed DBI metric is well correlated with the Mean Opinion Score obtained from subjective quality assessment~\cite{kapoor2021capturing}. While PPP has the lowest DBI score (0.005), Fig.~\ref{fig:deband_deepDeband} shows that the debanded image appears blurry with noticeable banding transitions. WaveMamba and its variants achieve low DBI scores, consistent with visual inspection, indicating the closest alignment with human perception. The WaveMamba variants demonstrate that the proposed frequency masking map leads to lower DBI scores and achieves high PSNR and SSIM, proving its effectiveness in improving the removal of banding artifacts while maintaining structural consistency.

Table~\ref{tab: cross_dataset} presents the cross-dataset evaluation on the BAND-2k dataset using models trained on the deepDeband test set. Although CAMBI is designed to quantify banding artifacts, it again exhibits limited correlation with perceptual quality. For example, the Banded Images yield the lowest CAMBI score (0.513), despite being the original input and displaying severe visual artifacts. In contrast, WaveMamba-WWM achieves the highest PSNR (39.055dB) and lowest LPIPS (0.059), while WaveMamba-MAP attains the lowest DBI score (0.082). This indicates that metrics such as CAMBI deviate from human visual perception, whereas DBI, built on a banding learning framework, aligns more closely with human visual judgment.

\subsection{Visual Inspection}
\begin{figure*}[htbp]
    \centering
    \centerline{\includegraphics[width=0.95\textwidth]{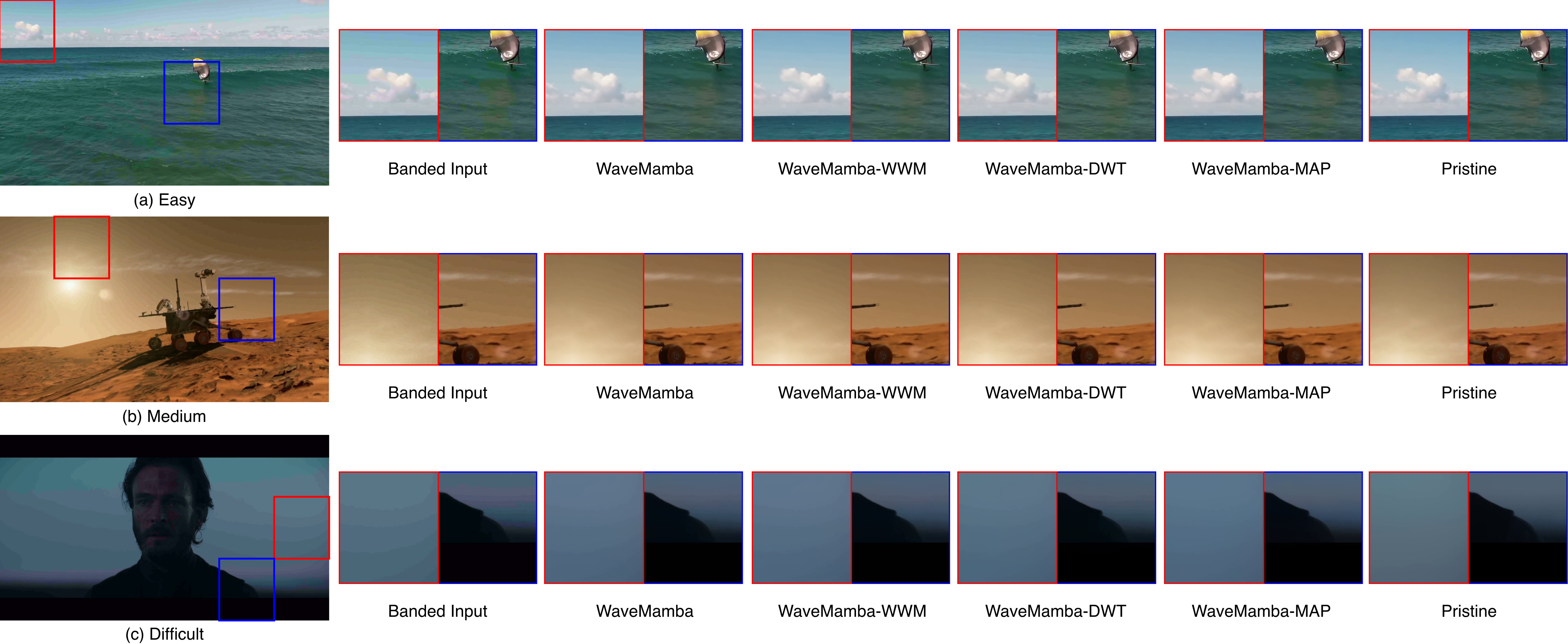}}
    \caption{Visual comparison of debanding performance during inference on the BAND-2k dataset. Each row represents a difficulty level: (a) easy, (b) medium, and (c) difficult based on DBI values. 
    Columns correspond to: Banded Input, WaveMamba variants, and Pristine.}
    \label{fig:deband_band-2k}
\end{figure*}

\begin{figure}[htbp]
    \centering
    \centerline{\includegraphics[width=0.5\textwidth]{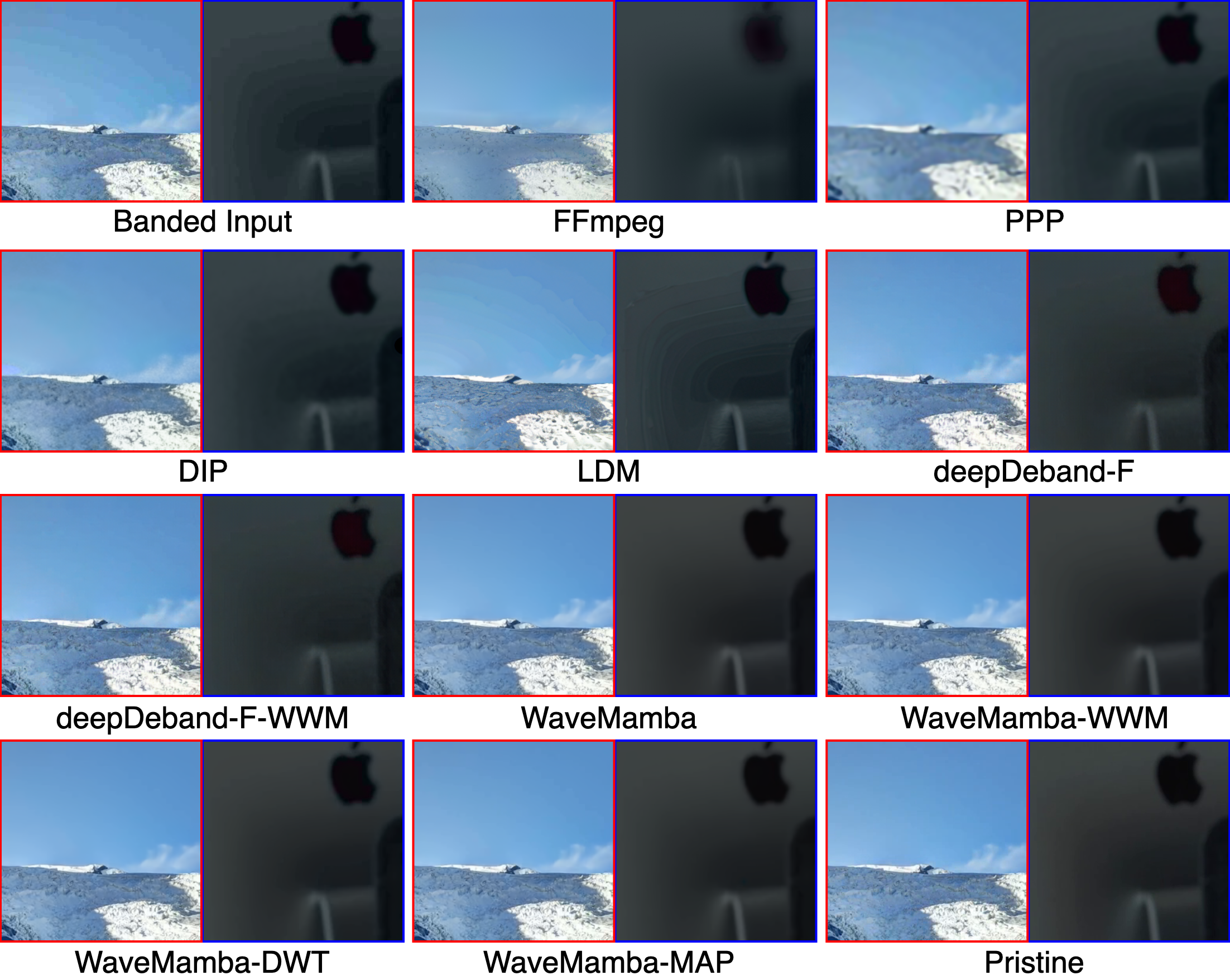}}
    \caption{Qualitative comparisons of the models trained and tested on deepDeband patches.}
    \label{fig:deband_deepDeband}
\end{figure}

The visual inspection reveals the weakness of the banding metrics. We selected three examples from BAND-2k based on ascending DBI scores, corresponding to easy, medium, and difficult cases (Fig.~\ref{fig:deband_band-2k}). Each example includes the banded input and selected zoomed-in patches. WaveMamba variants reduce banding artifacts across all difficulty levels, producing smoother transitions. Notably, WaveMamba-MAP achieves the best performance in detail preservation and perceptual quality close to pristine images, particularly in sky and low-luminance regions.
Besides these, we selected two output samples from the deepDeband test set. As shown in Fig.~\ref{fig:deband_deepDeband}, we can see that when performing debanding at a patch level, WaveMamba variants visually outperform all other methods. Compared to deepDeband variants, they provide smoother areas with fewer visible banding artifacts. However, banding metrics such as CAMBI and BBAND indicate that deepDeband achieves the lowest scores, which contradicts our visual inspection.

\section{Conclusion and Future Work}
\label{sec: con}
We proposed a banding artifact restoration method that leverages the WaveMamba network combined with a weighted wavelet map. By interpolating and normalizing high-frequency components from multi-scale wavelet decomposition, we effectively reduced banding artifacts while preserving textural structures. We constructed a banding restoration benchmark and evaluated the performance of current mainstream methods. Finally, the results revealed a misalignment between existing banding metrics and human perception, underscoring the need for more reliable and perceptually aligned banding quality metrics.


\bibliographystyle{IEEEtran}
\bibliography{refs, smaras}
\end{document}